\documentclass[twoside,11pt]{article}

\PassOptionsToPackage{sort&compress}{natbib}

\usepackage[preprint]{jmlr2e}

\usepackage{microtype}

\usepackage{multirow}
\usepackage{booktabs}

\usepackage{tikz}
\usetikzlibrary{positioning, arrows.meta, shapes.geometric, calc, fit, backgrounds}

\usepackage{pdfpages} 
\usepackage{lastpage} 

\newcommand{\code}[1]{\texttt{#1}}

\jmlrheading{Vol}{2026}{1-\pageref{LastPage}}{SUBMIT; Revised REVISE}{PUBLISH}{ID}{Colin Gaffney, Shutong Li, Daniel Ng, Anastasia Petrushkina, Niket Kumar, Adam Cogdell, Mridul Sahu, Yaning Liang, Nikhil Bansal, Justin Pan, Angel Mau, Abhishek Agrawal, Marco Berlot, Rakesh Iyer, Ruoxin Sang, Kiranbir Sodhia}

\ShortHeadings{Orbax: Distributed Checkpointing with JAX}{Gaffney, Li, Ng, et al.}
\firstpageno{1}

\begin{document}


\title{Orbax: Distributed Checkpointing with JAX}

\author{%
\noindent{\textnormal{\normalsize \textsc{Lead Authors}}}\hfill\\
\name Colin Gaffney$^{1}$, \name Shutong Li$^{1}$, \name Daniel Ng$^{1}$\hfill\\
\email \{cpgaffney, lishutong, dnlng\}@google.com\hfill\\[0.4cm]
\noindent{\textnormal{\normalsize \textsc{Contributors}}}\hfill\\
\name Anastasia Petrushkina$^{3}$, \name Niket Kumar$^{1}$, \name Adam Cogdell$^{1}$, \name Mridul Sahu$^{1}$, \name Yaning Liang$^{1}$, \name Nikhil Bansal$^{1}$, \name Justin Pan$^{1}$, \name Angel Mau$^{1}$, \name Abhishek Agrawal$^{1}$, \name Marco Berlot$^{3}$\hfill\\
\email \{dicentra, niketkb, adamcogdell, mridulsahu, yaning, nikhilbansall, justinpan, angelmau, abhisekar, mxberlot\}@google.com\hfill\\[0.4cm]
\noindent{\textnormal{\normalsize \textsc{Leadership}}}\hfill\\
\name Rakesh Iyer$^{1}$, \name Ruoxin Sang$^{2}$, \name Kiranbir Sodhia$^{2}$\hfill\\
\email \{rni, rxsang, ksodhia\}@google.com\hfill\\[0.2cm]
\addr $^1$Google, Mountain View, USA \\
      $^2$Google DeepMind, Mountain View, USA \\
      $^3$Google DeepMind, London, UK
}

\editor{My editor}

\maketitle

\begin{abstract}%
In a landscape of high-performance distributed ML systems, JAX has emerged as a framework of choice. However, JAX's modular design philosophy leaves it without a standardized checkpointing solution. In this paper, we introduce Orbax, a modular, JAX-native checkpointing library that abstracts the complexities of distributed accelerator systems while also providing flexibility for user-friendly checkpoint manipulations throughout the ML model lifecycle. We demonstrate performance exceeding comparable PyTorch competitors by up to 3.5$\times$ for saving and 2$\times$ for loading. The library is available at \url{https://github.com/google/orbax}.
\end{abstract}

\begin{keywords}
  Checkpoint, Checkpointing, JAX, Orbax
\end{keywords}

\section{Introduction}

The JAX \citep{jax2018github} library has gained prominence as a high-performance numerical computing framework for machine learning (ML). However, JAX’s design philosophy encourages a minimal, modular core, leaving domain-specific functionalities like checkpointing---the persistent representation of model states---fragmented or absent \citep{jaxstack2024}.

Checkpointing operations must be orchestrated carefully to allow for fast training recovery without idling expensive accelerator hardware \citep{cottier2025risingcoststrainingfrontier}. Checkpoints provide essential resilience against hardware failures and preemptions \citep{kokolis2025revisitingreliabilitylargescalemachine} but saving snapshots naturally disrupts computations. While traffic between accelerators and host memory is highly performant, transfers to remote persistent storage are bottlenecked by variable network and disk speeds. Strategies like asynchronous checkpointing \citep{nicolae2020deepfreeze} and dynamic checkpointing intervals \citep{mohan2021checkfreq, jin2025adaptivefaulttolerancemechanisms} are crucial for maximizing training efficiency \citep{geminiteam2025geminifamilyhighlycapable}. Recovery performance is also key to accelerating developer productivity and minimizing downtime. Checkpoints are often loaded across different accelerator topologies \citep{wan2025bytecheckpointunifiedcheckpointinglarge} using a technique called \textit{resharding}.

Beyond performance, modern ML development introduces usability challenges. Checkpoints require a unified format for consumption in pre-training, post-training, evaluation, and inference codebases. They require customization and flexibility in loading strategies to enable experimentation with techniques like LoRA \citep{hu2021loralowrankadaptationlarge}. They must also support single- and multi-controller workloads while seamlessly inter-operating with various cloud storage backends.

To address these challenges, we introduce Orbax, a modular JAX-focused checkpointing library built for high performance at cutting-edge scales. Orbax abstracts the complexities of distributed storage and inter-process coordination, minimizing resource costs while maximizing developer productivity.

\section{Related Work}

Prior efforts to support JAX workloads have generally suffered from limitations in scalability, generalizability, or usability. Flax \citep{flax2020github}, for instance, provided different APIs for checkpointing in single- and multi-host contexts, while failing to scale due to object size restrictions. Later solutions leveraged TensorStore \citep{tensorstore2020github}, but often at the wrong level of abstraction, requiring significant manual array manipulation and distributed systems interaction. Training frameworks like T5X \citep{roberts2022scalingmodelsdatatextttt5x} and Pax \citep{paxml2022github} introduced bespoke checkpointing libraries. Developers working with Pathways \citep{barham2022pathwaysasynchronousdistributeddataflow} relied on deeply integrated, closed-source checkpointing approaches.

In the PyTorch ecosystem, the Distributed Checkpoint (DCP) library \citep{pytorch_dcp_2022} provides an analogous checkpointing framework, prioritizing distributed I/O, asynchronous saving, and topology-agnostic formats. However, DCP is tailored for the core task of distributed array tree I/O, delegating step management to user training frameworks. NeMo \citep{kuchaiev2019nemo} provides similar core functionalities, also incorporating novel strategies like Fully Parallel Saving (FPS), which distributes save traffic across data-parallel ranks. While conceptually the innovations of these libraries can generalize, JAX users cannot benefit from them out-of-the-box.

\section{Design}

Orbax's API design (visualized in Appendix \ref{api_diagram}, Figure \ref{fig:api_diagram}) emphasizes usability through core support for the following dominant use cases:

\begin{itemize}
\item \textbf{Sequence-of-steps}: ML training use cases commonly perform computation using a sequence of steps, where progress may be saved every \code{n} steps. In case of interruption, the latest step can be used to restart training. Orbax defines \code{training.Checkpointer} to support step-level orchestration. This provides an object-oriented interface to record training properties (recent steps, ongoing saves), and to perform curation tasks (garbage collection, save interval selection).
\item \textbf{Singular checkpoints}: Users desiring more control and flexibility need not conform to the assumptions imposed by the sequence-of-steps use case. Instead, a checkpoint may be simply represented as a file path. Orbax defines a purely functional interface to facilitate more direct interaction with specific checkpoints, allowing for targeted usage and debugging.
\end{itemize}

Following the principle of progressive disclosure of complexity, Orbax then exposes users to layers designed for flexibility and advanced customization. These include the concept of a \textbf{checkpointable} (short for \textbf{checkpointable object}), a discrete, independently-serializable sub-component of the checkpoint as a whole, conforming to two primary principles:

\begin{itemize}
\item \textbf{Checkpointables are separable}: Checkpointable \code{A} may be loaded and used in some contexts without the need for checkpointable \code{B}, and vice versa. The checkpoint as a whole can still be useful even if a particular checkpointable is deleted or never saved in the first place. The optimizer state and dataset iterator states, for example, can typically be ignored for evaluations and discarded for inference.
\item \textbf{Checkpointables have distinct types}: Different pieces of the overall training state are often represented as logically distinct Python objects, which naturally lend themselves to different serialization logic and on-disk representations. For example, model parameters and optimizer states are represented as nested trees of large distributed arrays, while a dataset iterator can be represented in a checkpoint as a simple index. 
\end{itemize}

The checkpointable concept provides a natural layer for introducing user-defined customizations. Orbax provides a \code{CheckpointableHandler} interface exposing methods such as \code{save}, \code{load}, and \code{metadata}. Alternatively, the \code{StatefulCheckpointable} interface allows \code{save} and \code{load} methods to be implemented directly on the checkpointable object itself. This allows custom objects (e.g. Grain dataset iterators \citep{grain2023github}) to add Orbax checkpointing support without taking an explicit dependency, allowing for substantial modularity within the JAX ecosystem.

Orbax provides an abstraction that allows users to establish a well-defined contract between their mental image of a model's structure compared to its actual representation on disk. This increases reliability and comprehensibility as checkpoints are transferred between codebases, resharded at read-time onto different topologies, or partially loaded to exclude optimizers or other states. Orbax thus formalizes the concept of \textbf{checkpoint} vs. \textbf{abstract checkpoint} (equivalently \textbf{checkpointable} vs. \textbf{abstract checkpointable}). All concrete types define an abstract type, a lightweight representation of the object recording key properties (e.g. shapes, dtypes, shardings). The abstract type can then be used to describe the concrete object as metadata or to customize loading behavior (e.g. through reshaping, casting, or resharding).

\section{Performance}

Due to the lack of an equivalent JAX-based library, we evaluate Orbax's performance against PyTorch's Distributed Checkpoint (DCP) library \citep{pytorch_dcp_2022}. We conduct evaluations on NVIDIA A100 (40GB) GPUs using a single-region Google Cloud Storage (GCS) bucket as the storage backend with Llama 3.1 8B, 70B, and 405B models (4, 16, and 64 devices respectively). For the JAX-based benchmarks, we use MaxText \citep{maxtext}, an LLM library and reference implementation, to obtain JAX-compatible Llama models. For each model size, we perform a total of \code{n=20} save and load operations in isolation to eliminate sources of variance arising from running a full training loop.

As summarized in Table \ref{tab:performance_comparison}, the results demonstrate that Orbax's performance is competitive with that of PyTorch DCP, and typically exceeds it at larger scales, though particularly for small models, the checkpoint file format used by Orbax may impose additional write overhead.

\begin{table}[!htb]
\centering
\caption{Performance Comparison: PyTorch DCP vs. Orbax}
\vbox{\vspace{1ex}\small\noindent Values are reported as \textit{P50 [P90]}. Speedup values display \textit{Median [Tail]} factors.}
\begin{tabular}{llccc}
\toprule
Operation & Model Size & PyTorch & Orbax & Speedup \\
\midrule
\multirow{3}{*}{Save (Blocking)\textsuperscript{*}} 
 & 8B   & \textbf{1.9s [2.1s]}   & 3.9s [4.1s]   & 0.5$\times$ [0.5$\times$] \\
 & 70B  & \textbf{5.1s [5.6s]}   & 8.8s [9.3s]   & 0.6$\times$ [0.6$\times$] \\
 & 405B & \textbf{7.9s [8.7s]}   & 12.8s [13.9s] & 0.6$\times$ [0.6$\times$] \\
\midrule
\multirow{3}{*}{Save (Background)} 
 & 8B   & \textbf{14.4s} [86.7s] & 33.9s \textbf{[35.2s]} & 0.6$\times$ \textbf{[2.5$\times$]} \\
 & 70B  & 211.0s [303.4s] & \textbf{104.6s [135.4s]} & \textbf{2.0$\times$ [2.2$\times$]} \\
 & 405B & 595.0s [601.7s] & \textbf{173.0s [216.9s]} & \textbf{3.4$\times$ [2.8$\times$]} \\
\midrule
\multirow{3}{*}{Load} 
 & 8B   & 45.1s [48.7s]  & \textbf{22.6s [24.9s]}  & \textbf{2.0$\times$ [2.0$\times$]} \\
 & 70B  & 109.7s [123.9s] & \textbf{75.3s [84.5s]}  & \textbf{1.5$\times$ [1.5$\times$]} \\
 & 405B & 169.5s [177.6s] & \textbf{117.2s [125.1s]} & \textbf{1.4$\times$ [1.4$\times$]} \\
\bottomrule
\end{tabular}

\vspace{1ex}
\raggedright
{\footnotesize \textsuperscript{*} \textit{The blocking portion of the save is almost entirely spent copying arrays from device to host memory, for which Orbax relies on out-of-the-box JAX / XLA functionality. Blocking performance is thus only minimally attributable to Orbax itself.}\par}
\label{tab:performance_comparison}
\end{table}

Further performance evaluations can be found in the appendices, including for loading-with-resharding (Appendix \ref{ReshardingAndChunking}), on-disk representations (Appendix \ref{StorageFormat}), and scalable checkpointing for up to 2000 nodes (Appendix \ref{Scalability}).

\section{Conclusions and Future Work}

We have presented Orbax, a JAX-based library for scalable distributed checkpointing. The API formalizes core use cases (sequence-of-steps, singular checkpoints) and modes of interaction (\textit{checkpointables}, abstract vs. concrete types), while facilitating a degree of customization and modularity. We have demonstrated its strong out-of-the-box performance relative to existing solutions available for PyTorch.

In the future, we will continue to optimize performance, in particular by testing at scales of 10-100K nodes and by tuning behavior for various supported storage backends. We aim to eliminate legacy, difficult-to-use APIs while introducing new supported modes of interaction, including model surgery, which facilitates online load-time checkpoint transformation and manipulation.

\newpage

\appendix

\section{API Diagram}
\label{api_diagram}

\begin{figure}[!htb]
    \centering
    \resizebox{1.0\textwidth}{!}{
        \begin{tikzpicture}[
            node distance=1.2cm and 1cm,
            header/.style={font=\LARGE\sffamily\bfseries, align=center, minimum height=0.5cm},
            box_white/.style={rectangle, draw=black, thick, rounded corners, minimum height=1.1cm, align=center, fill=white, inner sep=3pt, font=\large\sffamily},
            box_blue/.style={rectangle, draw=blue!80, thick, rounded corners, minimum height=1.1cm, align=center, fill=blue!5, inner sep=3pt, font=\large\sffamily},
            code_box_blue/.style={box_blue, font=\large\ttfamily},
            split_node_base/.style={rectangle, draw=blue!80, thick, rounded corners, inner sep=0pt}
        ]
        
        \node[header] (h1) at (0, 20) {Sequence-of-checkpoints};
        \node[header] (h2) at (7, 20) {Single-checkpoint};
        \node[header] (h3) at (16, 20) {Checkpointable};
        \node[header] (h4) at (21, 20) {PyTree Leaf};
        
        \draw[dashed, thick, draw=black!60] (-3, 13) -- (24, 13);
        
        \node[code_box_blue, minimum width=6cm] (cp) at (0, 18) {\texttt{training.}\\\texttt{Checkpointer}};
    
        \node[box_blue, minimum width=4.25cm, minimum height=2.4cm] (save_left) at (7, 18) {\texttt{save}\\\texttt{(sync/async)}};
        \node[code_box_blue, minimum width=4.25cm, minimum height=1.2cm, anchor=north west] (save_right_bottom) at (save_left.east) {\texttt{\_checkpointables}};
    
        \node[box_blue, minimum width=4.25cm, minimum height=2.4cm] (load_left) at (7, 15) {\texttt{load}\\\texttt{(sync/async)}};
        \node[code_box_blue, minimum width=4.25cm, minimum height=1.2cm, anchor=north west] (load_right_bottom) at (load_left.east) {\texttt{\_checkpointables}};
        
        \node[box_white, minimum width=4.5cm, font=\large] (stateful) at (16, 11.5) {\texttt{Stateful}\\\texttt{Checkpointable}};
        \node[box_white, minimum width=4.5cm, minimum height=1.3cm] (cp_handler) [below=0.3cm of stateful] {\texttt{Checkpointable}\\\texttt{Handler}\\{}[\texttt{T}, \texttt{AbstractT}]};
        \node[code_box_blue, minimum width=4.5cm] (pt_handler) [below=0.3cm of cp_handler] {\texttt{PyTreeHandler}};
        \node[code_box_blue, minimum width=4.5cm] (js_handler) [below=0.3cm of pt_handler] {\texttt{JsonHandler}};
        \node[code_box_blue, minimum width=4.5cm] (pr_handler) [below=0.3cm of js_handler] {\texttt{ProtoHandler}};
        \node[code_box_blue, minimum width=4.5cm] (ar_handler) [below=0.3cm of pr_handler] {\texttt{ArrayHandler}};
    
        \node[box_white, minimum width=4.5cm, font=\large] (leaf_handler) at (21, 11.5) {\texttt{LeafHandler}\\{}[\texttt{L}, \texttt{AbstractL}]};
        \node[code_box_blue, minimum width=4.5cm] (array_handler) [below=0.3cm of leaf_handler] {\texttt{ArrayHandler}};
        \node[code_box_blue, minimum width=4.5cm] (numpy_handler) [below=0.3cm of array_handler] {\texttt{NumpyHandler}};
        \node[code_box_blue, minimum width=4.5cm] (scalar_handler) [below=0.3cm of numpy_handler] {\texttt{ScalarHandler}};
        \node[code_box_blue, minimum width=4.5cm] (string_handler) [below=0.3cm of scalar_handler] {\texttt{StringHandler}};
    
        \node[code_box_blue, minimum width=5cm, minimum height=1.1cm] (context) at (4.5, 11.5) {\texttt{Context}};
        
        \draw[<->, >=latex, line width=1.5pt, draw=black!70] (25, 9) -- (25, 16);
        
        \node[align=center, anchor=south, font=\large\sffamily\bfseries, text=black!80] at (25, 16.2) {Primary\\entry points};
        
        \node[align=center, anchor=north, font=\large\sffamily\bfseries, text=black!80] at (25, 8.8) {Customization\\layers};
        
        \end{tikzpicture}
    }
    \caption{Orbax Checkpoint API structure}
    \label{fig:api_diagram}
\end{figure}
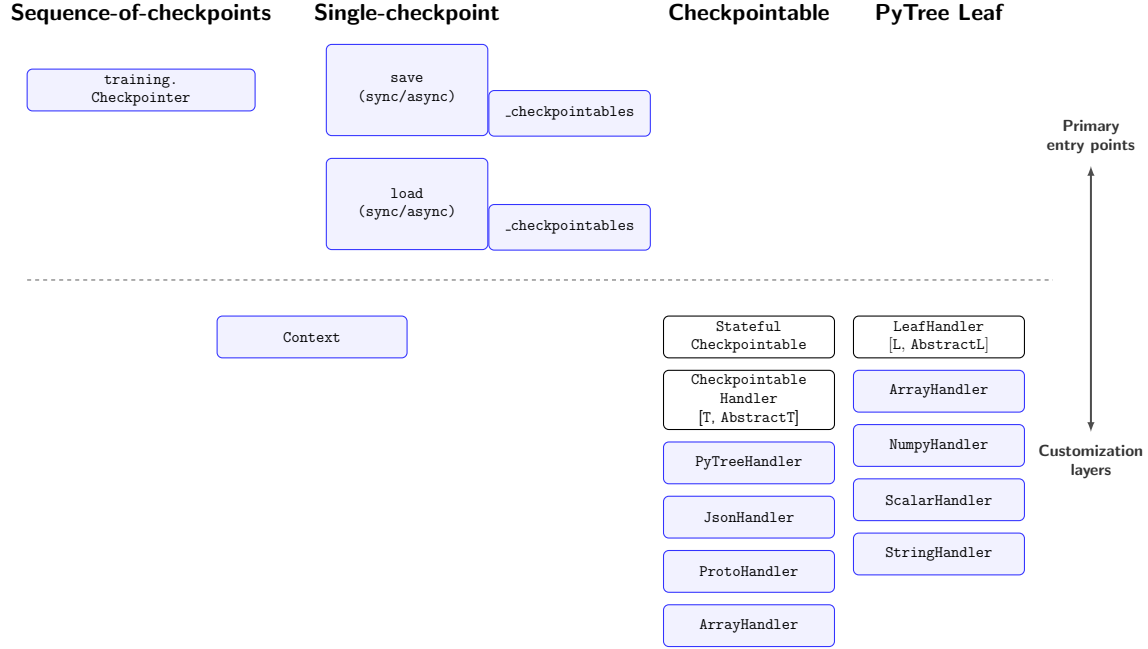

\section{Ecosystem Compatibility}

As we have noted above, fragmentation in the JAX ecosystem was an important motivator in the development of Orbax. Incompatibility of checkpoints across different codebases acted as a brake on development and experimentation. While Orbax provides a shared standard format and interface for JAX users, fragmentation between the JAX and PyTorch ecosystems remains a challenge.

While it is beyond our scope to provide a shared checkpoint format for both JAX and PyTorch, our core API facilitates loading formats like Safetensors \citep{safetensors2023hf}, a format commonly used for sharing open-source models. This is accomplished via the \code{CheckpointLayout} abstraction, an internal layer allowing Orbax to easily branch between different on-disk representations while reusing format-agnostic logic for managing checkpoints (e.g. async checkpointing logic, sequence-of-steps logic, etc.).

\begin{verbatim}
    from orbax.checkpoint import v1 as ocp
    
    with ocp.Context(
      checkpoint_layout=ocp.options.CheckpointLayout.SAFETENSORS
    )
    ocp.load("model.safetensors", abstract_state)
\end{verbatim}

\section{Saving Logic}

\textbf{Synchronous Phase:} When a checkpoint save is requested, certain tasks must be performed in a blocking fashion, while expensive operations like disk I/O should be moved into a background thread, so as to minimize the time spent blocking training on device. Orbax makes a design decision to perform certain validations (even some that require cheap I/O, like existence checks) in the synchronous phase so as to better catch and report common errors to the user. 

The more critical operation performed during the synchronous phase copies weights from accelerator device to host memory. This prevents concurrent modification of the weights while the checkpoint write is ongoing, and is required regardless for hardware that does not support direct write from accelerator memory to storage. 
When copying from device to host, each worker process identifies globally unique shards for each parameter on its local devices.

\begin{figure}[!htb]
\centering
\resizebox{\textwidth}{!}{
\begin{tikzpicture}[
    x=1.1cm, y=3.2cm,
    train/.style={draw=green!50!black, fill=green!10, thick, minimum width=2.2cm, minimum height=1.2cm, align=center, font=\Large},
    sync/.style={draw=red!60!black, fill=red!10, thick, minimum height=1.2cm, align=center, font=\Large},
    async/.style={draw=blue!60!black, fill=blue!10, thick, minimum height=1.2cm, align=center, font=\Large},
    disk/.style={cylinder, draw=black!70, fill=gray!20, thick, aspect=0.15, minimum height=2.5cm, minimum width=6.5cm, shape border rotate=90, align=center, font=\Large},
    >=Stealth
]

\node[anchor=east, font=\Large\sffamily\bfseries] at (-0.5, 3) {Accelerator};
\node[anchor=east, font=\Large\sffamily\bfseries] at (-0.5, 2) {Host CPU (Main)};
\node[anchor=east, font=\Large\sffamily\bfseries] at (-0.5, 1) {Host CPU (Background)};
\node[anchor=east, font=\Large\sffamily\bfseries] at (-0.5, 0) {Storage};

\draw[dotted, gray!60, thick] (-0.5, 2.5) -- (23, 2.5);
\draw[dotted, gray!60, thick] (-0.5, 1.5) -- (23, 1.5);
\draw[dotted, gray!60, thick] (-0.5, 0.5) -- (23, 0.5);

\node[train] (t1) at (1, 3) {Step $i-1$};
\node[train] (t2) at (3.5, 3) {Step $i$};
\node[train] (t3) at (14.5, 3) {Step $i+1$};
\node[train] (t4) at (17, 3) {Step $i+2$};
\node[train] (t5) at (19.5, 3) {Step $i+3$};
\node[train] (t6) at (22, 3) {Step $i+4$};

\node[sync, minimum width=2.2cm] (val) at (6.5, 2) {Validation};
\node[sync, minimum width=5cm] (d2h) at (10.5, 2) {Device-to-Host Copy};

\node[async, minimum width=3.2cm] (meta) at (14.5, 1) {Global Metadata \\ (Leader)};
\node[async, minimum width=4cm] (write) at (18.5, 1) {Distributed Writes \\ (All)};
\node[async, minimum width=2.5cm] (fin) at (22, 1) {Finalize \\ (Leader)};

\node[disk] (disk_main) at (18.5, 0) {Storage};

\draw[->, thick] (t2.east) -- ++(0.5,0) |- (val.west) node[near end, above, xshift=-1.5cm, font=\ttfamily\Large] {save()};

\draw[->, thick] (val.east) -- (d2h.west);

\draw[->, thick, dashed, draw=green!50!black] (d2h.east) -- ++(0.5,0) |- (t3.west) node[near start, right, font=\Large\itshape, yshift=-0.2cm] {Training Resumes};

\draw[->, thick] (d2h.south) |- (meta.west);

\draw[->, thick] (meta.east) -- (write.west);
\draw[->, thick] (write.east) -- (fin.west);

\draw[->, thick, dashed] (meta.south) -- (disk_main.north -| meta.south);
\draw[->, thick, dashed] (write.south) -- (disk_main.north);
\draw[->, thick, dashed] (fin.south) -- (disk_main.north -| fin.south);

\end{tikzpicture}

}
\caption{Asynchronous Saving}
\label{fig:orbax_save_timeline}
\end{figure}
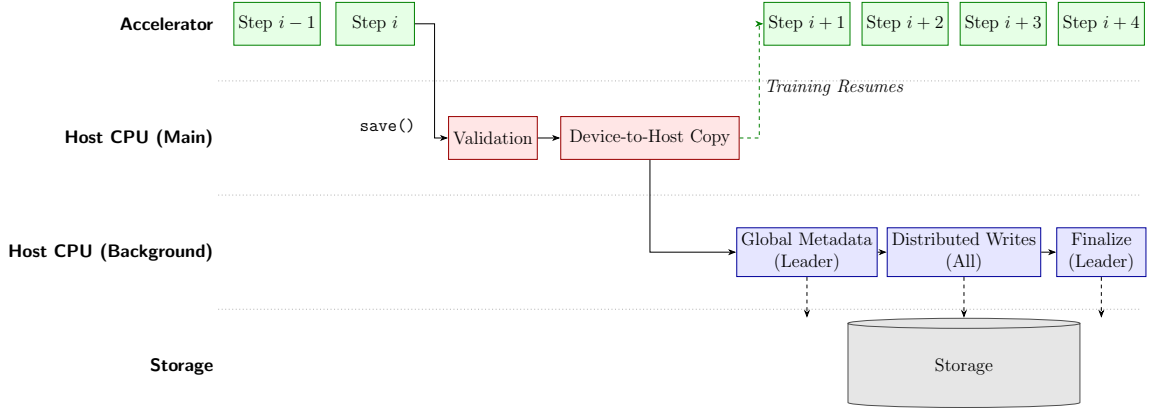

\textbf{Asynchronous Phase:} When an in-memory copy of the weights is obtained, control may return to the caller of \code{save} (if called using the asynchronous API) and a background thread to manage storage I/O may be started. See Figure \ref{fig:orbax_save_timeline} for a depiction of the synchronous / asynchronous saving timeline.

\textbf{Metadata Operations:} For a given checkpoint path, data is written to a temporary directory corresponding to the requested path. The temporary directory may or may not be the same as the final path, depending on whether the underlying filesystem supports atomic directory renames. This allows Orbax to ensure atomicity of the written checkpoint, which guards against interruptions while writing; if the temporary location is not finalized, the checkpoint cannot be considered complete.

Following the creation of the temporary location, further operations may begin. One background operation creates a metadata file describing the tree structure, used primarily to record key names, key types (\code{dict}, \code{list}, \code{tuple}, ...), and empty nodes. This information is considered to be global, shared by all worker processes, and is written uniquely by a single leader process. Additional metadata files may be written per-process, recording array shapes, shard shapes, and storage chunk shapes. Validation can be performed on these properties during finalization to ensure consistency across worker processes (see below). Orbax itself does not record all relevant metadata properties, but delegates to TensorStore for properties like data types.

\textbf{Distributed Writes:} A given host \code{X} creates a subdirectory of the same name, to which all of its local array data, previously transferred from local devices, may be written independently of other hosts, reducing the potential for scaling bottlenecks.

\begin{figure}[htbp]
    \centering
    \begin{tikzpicture}[
        font=\sffamily\small,
        box/.style={rectangle, minimum height=0.6cm, minimum width=2.4cm, align=center, thick},
        accbox/.style={box, draw=orange!80, fill=orange!5},       
        membox/.style={box, draw=green!60!black, fill=green!5},   
        pathbox/.style={draw, rectangle, minimum height=0.6cm, align=center, thick, outer sep=0pt},
        host/.style={draw, rectangle, inner sep=8pt, thick},
        arrow/.style={-{Stealth[scale=1.0]}, thick}
    ]

    \node[pathbox, minimum width=2.4cm] (db) at (0, 0) {/ckpt/global\_metadata};
    \node[pathbox, minimum width=2.8cm, anchor=west] (p0) at ([xshift=-\pgflinewidth]db.east) {/ckpt/n/process0};
    \node[pathbox, minimum width=2.8cm, anchor=west] (p1) at ([xshift=-\pgflinewidth]p0.east) {/ckpt/n/process1};
    \node[pathbox, minimum width=2.0cm, anchor=west] (pdots) at ([xshift=-\pgflinewidth]p1.east) {\dots};
    \node[pathbox, minimum width=2.8cm, anchor=west] (pn) at ([xshift=-\pgflinewidth]pdots.east) {/ckpt/n/process\_n};

    \begin{pgfonlayer}{background}
        \node [draw, thick, cylinder, shape border rotate=90, aspect=0.1, fit=(db) (pn), inner xsep=8pt, inner ysep=18pt, fill=gray!10] (storage) {};
    \end{pgfonlayer}
    \node [anchor=south, yshift=4pt] at (storage.south) {Storage};

    \def\hostY{4.0}
    \def\innerYspace{1.0}
    \def\topPadding{0.6} 

    \coordinate (H0_X) at (1.2, \hostY);
    \node[membox] (mem0) at (H0_X) {Host Memory};
    \node[accbox] (acc0) at ($(mem0.north) + (0, \innerYspace)$) {Accelerator 0 \\ HBM};
    \coordinate (top0) at ($(acc0.north) + (0, \topPadding)$); 
    \node[host, fit=(top0) (acc0) (mem0)] (h0_box) {};
    \node[anchor=north, yshift=-2pt] at (h0_box.north) {Host 0};

    \coordinate (H1_X) at ([yshift=\hostY cm]p1.center);
    \node[membox] (mem1) at (H1_X) {Host Memory};
    \node[accbox] (acc1) at ($(mem1.north) + (0, \innerYspace)$) {Accelerator 1 \\ HBM};
    \coordinate (top1) at ($(acc1.north) + (0, \topPadding)$); 
    \node[host, fit=(top1) (acc1) (mem1)] (h1_box) {};
    \node[anchor=north, yshift=-2pt] at (h1_box.north) {Host 1};

    \coordinate (HN_X) at ([yshift=\hostY cm]pn.center);
    \node[membox] (memN) at (HN_X) {Host Memory};
    \node[accbox] (accN) at ($(memN.north) + (0, \innerYspace)$) {Accelerator N \\ HBM};
    \coordinate (topN) at ($(accN.north) + (0, \topPadding)$); 
    \node[host, fit=(topN) (accN) (memN)] (hN_box) {};
    \node[anchor=north, yshift=-2pt] at (hN_box.north) {Host N};

    \node at ($(h1_box.east)!0.5!(hN_box.west)$) {\Large $\dots$};


    \draw[arrow] (acc0.south) -- (mem0.north);
    \draw[arrow] (acc1.south) -- (mem1.north);
    \draw[arrow] (accN.south) -- (memN.north);


    \draw[arrow, draw=red!60, densely dashed] (h0_box.south) -- (db.north) node[pos=0.25, xshift=-05pt, fill=white, inner sep=2pt, font=\scriptsize, text=red!60] {Finalize};
    
    \draw[arrow] (h0_box.south) -- (p0.north);
    
    \draw[arrow] (h1_box.south) -- (p1.north);

    \draw[arrow] (hN_box.south) -- (pn.north);

    \end{tikzpicture}
    
    \caption{Distributed Saving}
    \label{fig:distributed_write}
\end{figure}
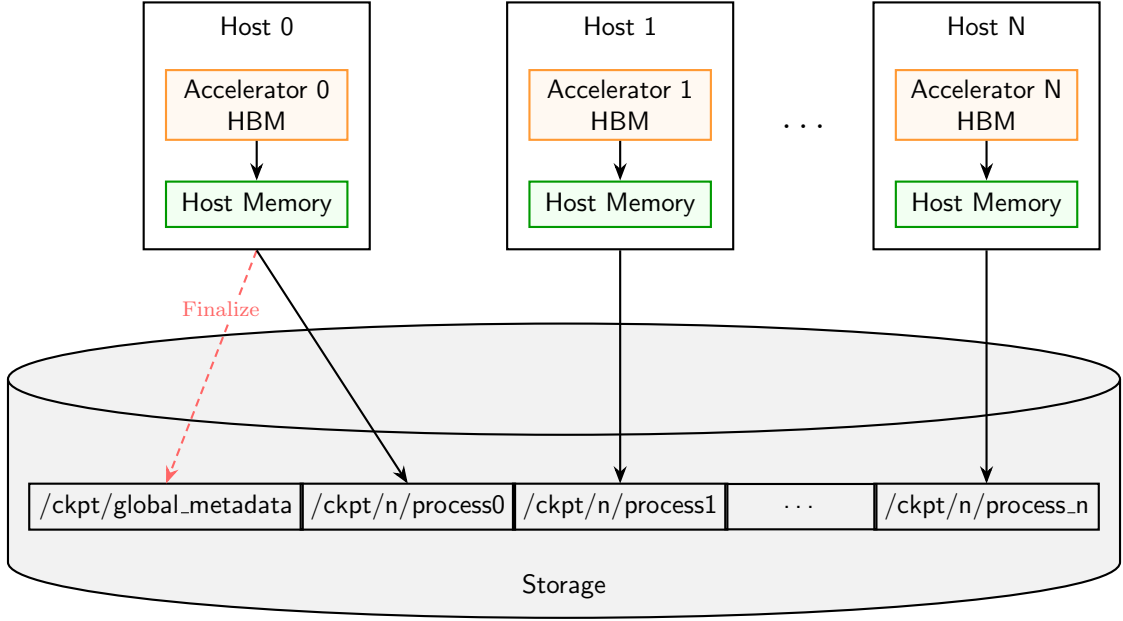

Once all processes have finished writing, a selected primary host performs a `finalize' operation to obtain a global view of the stored chunks, which were previously stored on a per-process basis. This finalization step reads per-host sharded array metadata and writes a merged index at the top directory level, which references the original data without performing any expensive deep copies (see Figure \ref{fig:distributed_write}). The global view facilitates checkpoint restoration on altered topologies, as we will further discuss later.

While a selected primary host always performs the final metadata merge, its exact identity is implementation-dependent, which will be discussed in greater detail in Section \ref{mcJaxVsPathways}. In multi-controller JAX (mcJAX), this role is filled by a designated leader process (typically process 0) which coordinates with followers via distributed barriers. In Pathways, the single controller naturally assumes this responsibility, orchestrating the merge once all remote worker I/O operations have been acknowledged.

Only when all operations are completed will the checkpoint be marked as `finalized', using either an indicator file or an atomic rename.

\section{Loading Logic}

Unlike saving, checkpoint loading is a much less frequent operation, and is thus less of a performance bottleneck for most applications (assuming asynchronous saving is used). In most training codebases, loading occurs synchronously after model compilation; the compiled model is used as a reference, or abstract, state needed to guide restoration (particularly by specifying sharding information for each weight) and validate the structure of the restored checkpoint. 

\textbf{Loading Plan:} The abstract state dictates how the checkpoint should be loaded, and is used by Orbax to construct a loading plan. As mentioned above, the abstract state is a nested tree structure with leaves of abstract arrays. The actual type of the abstract array corresponds to a concrete array type, into which the underlying data will be returned. This allows for dynamic casting between compatible object types; for example, between single-element numpy arrays and scalars or numpy arrays and JAX arrays. Other abstract array properties may include shape, which is commonly used for validation but also for truncating and padding for arrays with uneven sharding; dtype, which can be used for dynamic casting and precision reduction; and sharding, which describes the topology and partitioning onto which the array should be loaded, independent of its origin sharding.

\begin{figure}[!htb]
\centering
\begin{tikzpicture}[
    base/.style={align=center, minimum height=0.8cm, font=\sffamily\footnotesize, thick},
    decision/.style={base, diamond, aspect=2, draw=orange!80, fill=orange!5, inner sep=1pt},
    process/.style={base, rectangle, rounded corners, draw=blue!80, fill=blue!5, minimum width=3cm},
    data/.style={base, cylinder, shape border rotate=90, aspect=0.2, draw=gray!70, fill=gray!10, minimum width=2.2cm, minimum height=1cm},
    output/.style={base, rectangle, rounded corners, draw=green!60!black, fill=green!5, minimum width=3cm, font=\sffamily\bfseries},
    arrow/.style={-Stealth, thick, draw=black!80},
    data_arrow/.style={-Stealth, thick, draw=gray!60, densely dashed}
]

    \node[decision] (decide) at (0, 3.5) {Abstract State \\ Provided?};

    \node[data] (meta) at (0, 1.5) {Checkpoint \\ Metadata};

    \node[process] (use_meta) at (-3.8, 1.5) {Use Metadata \\ Properties};
    \node[process] (val_topo) at (-3.8, -0.5) {Validate Current \\ Topology};

    \node[process] (val_struct) at (3.8, 1.5) {Validate Structure};
    \node[process] (apply_prop) at (3.8, -0.5) {Apply Specified Properties \\ (Casting, Sharding, \\ Transformation, etc.)};

    \node[output] (plan) at (0, -2.5) {Loading Plan};

    
    \draw[arrow] (decide.west) -| node[near start, above, font=\scriptsize, xshift=-10pt] {No} (use_meta.north);
    \draw[arrow] (decide.east) -| node[near start, above, font=\scriptsize, xshift=10pt] {Yes} (val_struct.north);

    \draw[arrow] (use_meta.south) -- (val_topo.north);
    \draw[arrow] (val_struct.south) -- (apply_prop.north);

    \draw[arrow] (val_topo.south) |- (plan.west);
    \draw[arrow] (apply_prop.south) |- (plan.east);

    \draw[data_arrow] (meta.west) -- node[above, font=\scriptsize, text=gray!80] {reads} (use_meta.east);
    \draw[data_arrow] (meta.east) -- node[above, font=\scriptsize, text=gray!80] {reads} (val_struct.west);

\end{tikzpicture}
\caption{Generating the Loading Plan. When a user provides an abstract state, Orbax uses it as a strict contract to validate the nested structure and apply transformations. Otherwise, it relies on checkpoint metadata to validate the active accelerator topology.}
\label{fig:loading_plan_logic}
\end{figure}
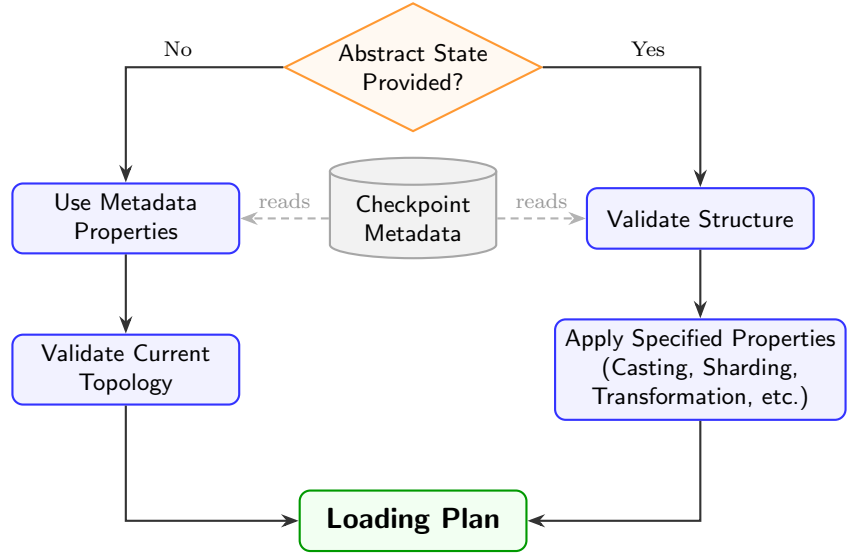

The API does not require an abstract tree; if not specified, metadata from the checkpoint itself will be used to obtain the necessary properties, as detailed in the logic flow of Figure \ref{fig:loading_plan_logic}. In this case, shardings of weights in the checkpoint must be validated against the current topology to ensure no discrepancies. If the device set has changed, Orbax does not attempt to infer a sharding on the user's behalf. Regardless, the metadata will be loaded in either case in order to validate the nested tree structure.

In the end, the abstract state is useful for establishing a contract between the user and the loading API, since the user may expect the object returned by \code{load} to have the same nested structure and properties as the abstract state. Ordinarily, it is assumed that any structural mismatch between the user-provided abstract state and the checkpoint metadata is evidence of a user error. However, the user may opt into a special \textit{partial loading} mode, in which the two structures may mismatch. If the abstract state specifies a subset of the keys contained in the checkpoint, only that subset will be loaded, and remaining keys will be ignored. Conversely, if the abstract state specifies a super-set of keys in the checkpoint, any keys missing from the checkpoint but present in the abstract state will be filled in with a placeholder value (in Python, \code{...}) to be replaced by concrete values by the user at a later point.

\textbf{Distributed Loading:} Given the user-specified loading properties, the process of loading shards from storage to host, then from host to device, is relatively straightforward. The sharding property dictates a specific assignment of global array index to host. Each host independently reads the data for the global array indices needed for its local devices. Logically, this operation is topology-agnostic, since it need only refer to the merged global metadata to identify the underlying data location. 

Once shards are loaded into host memory, they may be placed on device and a global view of the array constructed, without requiring data movement between devices. Global array construction naturally acts as a synchronization point and no individual process can return before all shards for all arrays have been materialized.

\section{Single- and Multi-Controller Checkpointing}
\label{mcJaxVsPathways}

JAX supports two modes of distributed coordination: single-controller (Pathways) and multi-controller (mcJAX). Orbax natively supports both cases and strives to hide all implementation details from the user. As long as the user has the necessary Pathways dependencies linked, all logic automatically switches over to single-controller checkpointing.

\textbf{Multi-Controller Checkpointing and Limitations:} The multi-controller programming model offered by mcJAX suffers from key checkpointing-related limitations; multi-process coordination, in particular, creates significant logic complexities. Specifically, a single leader process must be chosen to perform unique tasks, including directory creation, global metadata writing, per-process metadata merging, atomicity, and others. We also prefer to perform some operations by a unique host for performance reasons; for example, simple directory listing can introduce excessive load on the storage when O(1000) processes are concurrently attempting to retrieve the checkpoint history after recovering from an interruption (especially if the checkpoint history is long). In such cases, we prefer to interact with the filesystem only on the leader host and use XLA collectives to broadcast data to the remaining hosts. The leader-follower model in a multi-controller setting of course demands the use of barriers to prevent race conditions, which can often lead to subtle bugs. For example, all controllers need to check for a pre-existing directory on save. This validation must be performed in strict ordering with a subsequent directory creation, else some hosts may mistakenly identify a newly-created directory as belonging to a previous checkpoint. 

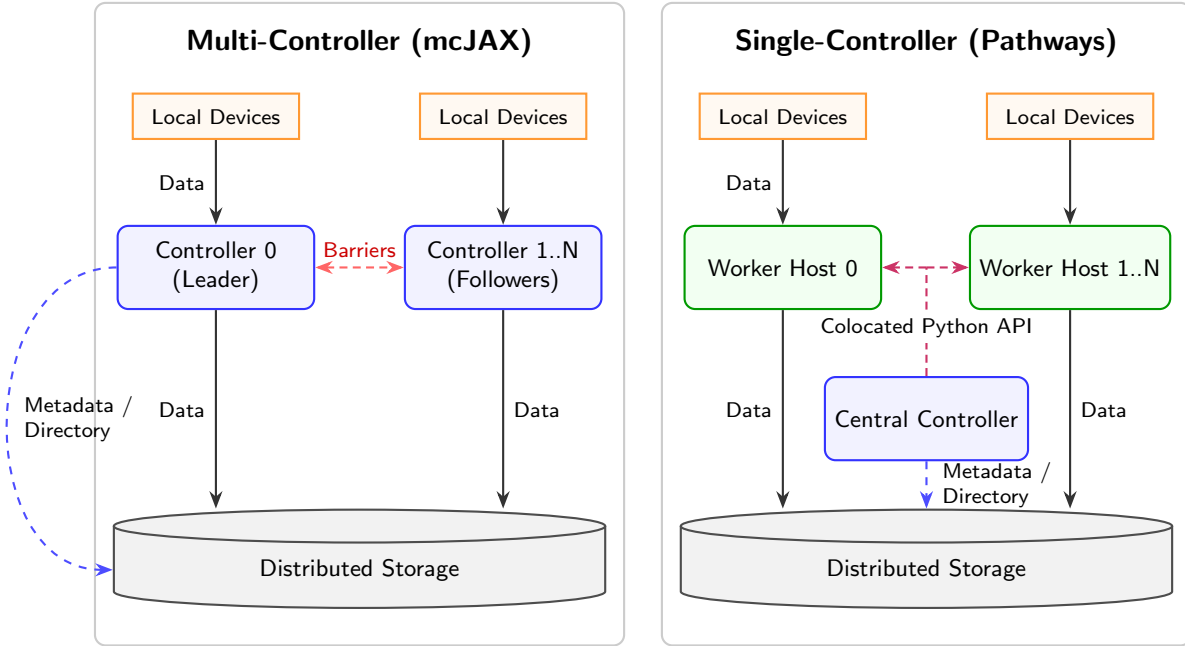
\begin{figure}[!htb]
\centering
\begin{tikzpicture}[
    node distance=1.5cm,
    ctrl/.style={rectangle, draw=blue!80, thick, rounded corners, fill=blue!5, align=center, minimum width=2.6cm, minimum height=1.1cm, font=\sffamily\footnotesize},
    worker/.style={rectangle, draw=green!60!black, thick, rounded corners, fill=green!5, align=center, minimum width=2.6cm, minimum height=1.1cm, font=\sffamily\footnotesize},
    device/.style={rectangle, draw=orange!80, thick, fill=orange!5, align=center, minimum width=2.2cm, minimum height=0.6cm, font=\sffamily\scriptsize},
    storage/.style={cylinder, shape border rotate=90, draw=black!70, thick, aspect=0.15, fill=gray!10, align=center, minimum width=6.5cm, minimum height=1.3cm, font=\sffamily\footnotesize},
    data_arrow/.style={-Stealth, thick, draw=black!80},
    meta_arrow/.style={-Stealth, thick, draw=blue!70, dashed},
    coord_arrow/.style={Stealth-Stealth, thick, draw=red!60, densely dashed},
    rpc_arrow/.style={-Stealth, thick, draw=purple!80, densely dashed},
    panel_box/.style={rectangle, draw=gray!40, thick, rounded corners, inner sep=0pt}
]

\useasboundingbox (0, -1.0) rectangle (14.5, 7.5);

\draw[panel_box] (0, -1.0) rectangle (7.0, 7.5);
\node[anchor=north, font=\sffamily\bfseries] at (3.5, 7.3) {Multi-Controller (mcJAX)};

\node[storage] (storage_mc) at (3.5, 0) {Distributed Storage};
\node[ctrl] (ctrl_leader) at (1.6, 4.0) {Controller 0\\(Leader)};
\node[ctrl] (ctrl_follower) at (5.4, 4.0) {Controller 1..N\\(Followers)};
\node[device] (dev_l) at (1.6, 6.0) {Local Devices};
\node[device] (dev_f) at (5.4, 6.0) {Local Devices};

\draw[data_arrow] (dev_l.south) -- node[left, font=\sffamily\scriptsize] {Data} (ctrl_leader.north);
\draw[data_arrow] (dev_f.south) -- (ctrl_follower.north);

\draw[data_arrow] (ctrl_leader.south) -- node[left, font=\sffamily\scriptsize] {Data} (ctrl_leader.south |- storage_mc.north);
\draw[data_arrow] (ctrl_follower.south) -- node[right, font=\sffamily\scriptsize] {Data} (ctrl_follower.south |- storage_mc.north);

\draw[coord_arrow] (ctrl_leader.east) -- node[above, font=\sffamily\scriptsize, text=red!80!black] {Barriers} (ctrl_follower.west);
\draw[meta_arrow] (ctrl_leader.west) to[out=180, in=180, looseness=1.2] node[right=2pt, font=\sffamily\scriptsize, align=left] {Metadata /\\Directory} (storage_mc.west);

\draw[panel_box] (7.5, -1.0) rectangle (14.5, 7.5);
\node[anchor=north, font=\sffamily\bfseries] at (11.0, 7.3) {Single-Controller (Pathways)};

\node[storage] (storage_pw) at (11.0, 0) {Distributed Storage};
\node[worker] (worker_0) at (9.1, 4.0) {Worker Host 0};
\node[worker] (worker_n) at (12.9, 4.0) {Worker Host 1..N};
\node[device] (dev_w0) at (9.1, 6.0) {Local Devices};
\node[device] (dev_wn) at (12.9, 6.0) {Local Devices};
\node[ctrl] (ctrl_central) at (11.0, 2.0) {Central Controller};

\draw[thick, draw=purple!80, densely dashed] (ctrl_central.north) -- (11.0, 4.0);
\draw[rpc_arrow] (11.0, 4.0) -- (worker_0.east);
\draw[rpc_arrow] (11.0, 4.0) -- (worker_n.west);
\node[font=\sffamily\scriptsize, align=center, fill=white, inner sep=2pt] at (11.0, 3.2) {Colocated Python API};

\draw[data_arrow] (dev_w0.south) -- node[left, font=\sffamily\scriptsize] {Data} (worker_0.north);
\draw[data_arrow] (dev_wn.south) -- (worker_n.north);

\draw[data_arrow] (worker_0.south) -- node[left, font=\sffamily\scriptsize] {Data} (worker_0.south |- storage_pw.north);
\draw[data_arrow] (worker_n.south) -- node[right, font=\sffamily\scriptsize] {Data} (worker_n.south |- storage_pw.north);

\draw[meta_arrow] (ctrl_central.south) -- node[right=2pt, font=\sffamily\scriptsize, align=left] {Metadata /\\Directory} (ctrl_central.south |- storage_pw.north);

\end{tikzpicture}
\caption{Architectural comparison of checkpointing execution under multi-controller (mcJAX) and single-controller (Pathways) environments. Pathways delegates high-bandwidth data transfers to worker hosts via the colocated Python API, eliminating the central controller as an I/O bottleneck.}
\label{fig:controller_architecture}
\end{figure}

\textbf{Single-Controller Checkpointing with Pathways:} To address these problems, we can leverage Pathways. JAX provides the \textit{Colocated Python} API, which grants \textit{`a uniform way to run Python code on the hosts associated with a set of JAX devices'} \citep{jax2024colocated}. This capability is integral to single-controller checkpoint, in which we must transfer data directly from devices to hosts to storage (and vice versa), without transferring data to the controller, which would introduce a significant bottleneck, illustrated in Figure \ref{fig:controller_architecture}.

Most checkpointing logic executes on controller host(s) in the same way regardless of the number of controllers, including API interactions, basic business logic, directory and metadata manipulation, etc. Indeed, for many data types (Numpy arrays, strings, scalars, etc.) that are already present in host memory and are not sharded across hosts or devices, serialization logic remains the same in mcJAX and Pathways. For on-device arrays, Orbax implements a different \code{LeafHandler}, which abstracts serialization logic for batches of tree leaves. Device/host transfers and read/write operations can then be scheduled using colocated Python directly on the remote worker hosts. The controller continues to take responsibility for core tasks like metadata writing as well as reading and writing data for Python objects (Numpy arrays, strings, scalars) stored by the controller itself.

The single-controller setting has many further advantages beyond our current scope, but it also offers greater user-friendliness and reduced debugging complexity. For example, it is easy to work with large checkpoints in a multi-worker setting (e.g. in a Jupyter notebook or Google Colab) as the notebook acts as a controller and connects to a large slice of devices, facilitating hands-on interaction with large checkpoints while requiring minimal setup.

\section{Additional Performance Features}

\subsection{Storage Format}
\label{StorageFormat}

Using TensorStore's \code{zarr} driver alone, Orbax supports a file format that represents every weight as a separate subdirectory, with each shard further represented as a distinct file. This can result in file size overhead and poor performance when there are many small weights, especially for models with unstacked layers. However, it is also low-overhead, so toy- and small-scale models can often realize better saving and loading performance with this format. It is also more human-readable, as individual parameters and their shardings can be easily inspected visually.

Using TensorStore's \code{zarr} driver layered on top of the \code{ocdbt} driver, the OCDBT (optionally-cooperative distributed B-tree) is Orbax's recommended default. This driver coalesces data for multiple array shards (potentially across multiple arrays) into single files of larger size. This enables easy tuning of file sizes for better performance across diverse filesystems. 

For OCDBT checkpoints, each in-memory array shard is coalesced with other concurrently written shards to be written as corresponding chunks in storage. Each chunk is described by a key in the OCDBT database, which represents all keys using a B+ tree for fast lookup of individual chunks. The OCDBT database also stores an array metadata key (\code{.zarray}) to record necessary properties. While the OCDBT format natively supports concurrent reads and writes onto shared files, doing so efficiently requires a coordinator server on a leader process. As noted above, this approach does not scale for hundreds or thousands of nodes. As such, Orbax isolates per-process writes, effectively performing non-cooperative writes on per-process subdirectories, with coordination only required at the end of the save operation.

Orbax's two supported checkpoint formats (via TensorStore) have differing performance characteristics demonstrating advantages for training setups of varying sizes. For the OCDBT-based format, we expect improved I/O speed for large-scale models, since the OCDBT driver can more efficiently group weights into files close to a target of 2 GiB. This format incurs more overhead at small scales though, as additional time is spent by TensorStore maintaining the B+ tree structure and by Orbax performing unnecessary metadata merges on a single process sub-directory.

Our evaluations use Llama 3.1 8B, 70B, and 405B on TPU topologies v5p-8, v5p-32, and v5p-128 (4, 16, and 64 devices) respectively. Seen in Table \ref{tab:ocdbt_performance}, the benefits of OCDBT only become clear with larger models and topologies.

\begin{table}[!htb]
\centering
\caption{Mean checkpoint load and save times (seconds) comparing OCDBT to a non-OCDBT baseline.}
\begin{tabular}{llrrr}
\toprule
Operation & Model & OCDBT & No OCDBT & Speedup \\
\midrule
\multirow{3}{*}{Load} 
 & 8B   & $22.85 \pm 4.68$ & $17.97 \pm 4.22$  & 0.79$\times$\textsuperscript{*} \\
 & 70B  & $41.19 \pm 4.17$ & $40.04 \pm 4.61$  & 0.97$\times$ \\
 & 405B & $59.14 \pm 4.06$ & $65.18 \pm 12.01$ & \textbf{1.10$\times$} \\
\midrule
\multirow{3}{*}{Save (Background)} 
 & 8B   & $59.37 \pm 9.04$  & $33.71 \pm 3.20$   & 0.57$\times$\textsuperscript{*} \\
 & 70B  & $62.47 \pm 9.09$  & $80.84 \pm 8.82$   & \textbf{1.29$\times$} \\
 & 405B & $91.76 \pm 14.91$ & $123.33 \pm 19.27$ & \textbf{1.34$\times$} \\
\bottomrule
\end{tabular}

\vspace{1ex}
\raggedright
\footnotesize{\textsuperscript{*} \textit{Indicates a performance regression where OCDBT overhead outweighs its scaling benefits at smaller model sizes.}}
\label{tab:ocdbt_performance}
\end{table}

\subsection{Resharding and Chunking}
\label{ReshardingAndChunking}

Orbax uses TensorStore's \code{zarr} driver, which is a chunked storage driver, meaning that it stores data as a rectangular n-dimensional grid of write chunks. In the case of distributed arrays, this corresponds neatly to the subdivision of a global array into a grid of per-device shards. Write operations for each shard must align to the write chunk grid. TensorStore also allows for a grid of read chunks, which may or may not be the same as the write chunks. For the \code{zarr} driver, read chunks must subdivide write chunks evenly (write chunk boundaries must align with some read chunk boundaries). Often referred to by Orbax as \textit{subchunks}, these represent the smallest readable unit, which we will discuss further below. At write time, read chunk boundaries must be configured statically, known in Orbax as \textit{subchunking}.

When loading, assuming the array's sharding has not changed since it was saved, the read range will correspond exactly write chunk in the database. If resharding is required, the read range may not align with write chunk boundaries. If the new shard aligns to the read chunk range (already specified during save as a static property), individual shard reads will again result in one chunk loaded per read request. However, if subchunking was not configured at write time, or if a read range happens to span multiple read chunks, some amount of data must be loaded and discarded as overhead.

We can consider how loading performance benefits from different chunked representations. An array in the pretrained checkpoint may have shape \code{(x, y)}, partitioned over 64 devices using \code{(16, 4)} partitions, giving shard shapes of (x / 16, y / 4). On a topology of 16 devices, using \code{(16, 1)} partitions, our shard shape is now \code{(x / 16, y)} instead of \code{(x / 16, y / 4)}. If we read a single shard on the 16-device topology, the ranges of the four source chunks perfectly subdivide our target shard. 

However, if we keep the number of devices constant but instead change the sharding, using \code{(64, 1)} partitions, we have shards of shape \code{(x / 64, y)}. Now, for each loaded shard, the same four source chunks must be loaded, but the range \code{x / 64 : x / 16} (in the row dimension) is discarded after loading, resulting in excess memory pressure and poor read performance.

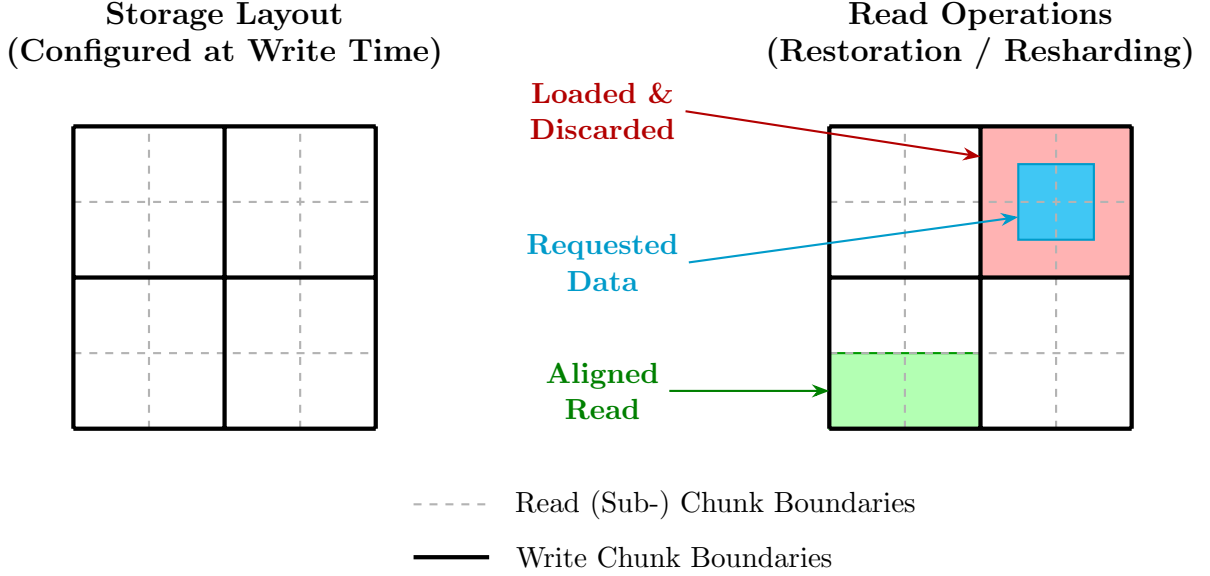
\begin{figure}[!htb]
\centering
\begin{tikzpicture}[
    node distance=2cm,
    write_grid/.style={ultra thick, draw=black},
    read_grid/.style={thick, dashed, draw=gray!60},
    overhead/.style={fill=red!30},
    requested/.style={fill=cyan!60, draw=cyan!80!black, thick},
    aligned/.style={fill=green!30, draw=green!60!black, thick},
    label_arrow/.style={->, >=Stealth, thick}
]

\begin{scope}[shift={(0,0)}]
    \node[font=\large\bfseries, align=center] at (2, 5.2) {Storage Layout\\(Configured at Write Time)};
    
    \fill[white] (0,0) rectangle (4,4);
    
    \draw[read_grid, step=1cm] (0,0) grid (4,4);
    
    \draw[write_grid, step=2cm] (0,0) grid (4,4);
\end{scope}

\begin{scope}[shift={(10,0)}]
    \node[font=\large\bfseries, align=center] at (2, 5.2) {Read Operations\\(Restoration / Resharding)};
    
    \fill[white] (0,0) rectangle (4,4);
    
    \draw[overhead] (2,2) rectangle (4,4);
    \draw[requested] (2.5,2.5) rectangle (3.5,3.5);
    
    \draw[aligned] (0,0) rectangle (2,1);
    
    \draw[read_grid, step=1cm] (0,0) grid (4,4);
    \draw[write_grid, step=2cm] (0,0) grid (4,4);
\end{scope}


\node[red!70!black, font=\bfseries, align=center] (label_ovh) at (7, 4.2) {Loaded \&\\Discarded};
\draw[label_arrow, red!70!black] (label_ovh.east) -- (12, 3.6);

\node[cyan!80!black, font=\bfseries, align=center] (label_req) at (7, 2.2) {Requested\\Data};
\draw[label_arrow, cyan!80!black] (label_req.east) -- (12.5, 2.8);

\node[green!50!black, font=\bfseries, align=center] (label_alg) at (7, 0.5) {Aligned\\Read};
\draw[label_arrow, green!50!black] (label_alg.east) -- (10, 0.5);

\begin{scope}[shift={(5,-0.5)}]
    \draw[read_grid] (-0.5, -0.5) -- (0.5, -0.5) node[right=0.2cm, black, font=\normalsize] {Read (Sub-) Chunk Boundaries};
    
    \draw[write_grid] (-0.5, -1.2) -- (0.5, -1.2) node[right=0.2cm, black, font=\normalsize] {Write Chunk Boundaries};
\end{scope}

\end{tikzpicture}
\vspace{0.5cm}
\caption{Visualizing aligned vs. unaligned read requests during checkpoint restoration.}
\label{fig:subchunking}
\end{figure}

If we subdivide the chunk grid used to store the checkpoint, we can obtain checkpoint chunk shapes of \code{(x / 64, y / 4)}, even though the shard shape remains \code{(x / 16, y / 4)}. When loading with \code{(64, 1)} partitions, the source chunks again perfectly subdivide the target shard, resulting in no overhead. This example is also illustrated visually by Figure \ref{fig:subchunking}.

We can demonstrate the performance of resharding under various scenarios using Llama 3.1 models (8B and 70B). Our first consideration is to obtain `source' (original) and `target' (new) shardings for our model. We rely on MaxText to generate model descriptors and shardings for a given topology. We then use these descriptors to load the models and immediately save them, using the desired configuration. At this point, the model checkpoint is sharded according to the `source' shardings, and may use either storage chunks aligned to shard boundaries or chunks subdividing shards. Chunk shapes are selected using an algorithm that attempts to limit the chunk size in bytes to a specified target (32 MiB for our experiments).

Our evaluations (see Table \ref{tab:resharding_performance}) measure performance when loading both the normal and subchunked checkpoints onto a variety of target shardings and topologies. Motivating use cases may be conceptualized as post-training a pre-trained checkpoint, resuming training with fewer chips, or resuming training on more memory-constrained hardware.

\begin{table}[!htb]
\centering
\caption{Mean checkpoint load time (seconds) during Resharding. Topologies are denoted using \code{data-fsdp-tensor} parallelism axes, where the number of devices is the product of these values.}
\begin{tabular}{llrrr}
\toprule
Model & Topology Shift & Subchunked & Not Subchunked & Speedup \\
\midrule
\multirow{3}{*}{70B} 
 & 4-8-4 $\rightarrow$ 1-128-1 & $13.41 \pm 1.44$ & $45.23 \pm 2.69$ & \textbf{3.37$\times$} \\
 & 4-8-4 $\rightarrow$ 1-32-1  & $14.40 \pm 0.51$ & $34.40 \pm 2.84$ & \textbf{2.39$\times$} \\
 & 4-8-4 (No change)           & \multicolumn{2}{c}{$15.99 \pm 1.50$} & -- \\
\midrule
\multirow{3}{*}{8B}  
 & 8-2-1 $\rightarrow$ 1-16-1  & $3.59 \pm 0.17$  & $20.40 \pm 2.56$ & \textbf{5.68$\times$} \\
 & 8-2-1 $\rightarrow$ 1-4-1   & $10.88 \pm 0.49$ & $21.07 \pm 3.12$ & \textbf{1.94$\times$} \\
 & 8-2-1 (No change)           & \multicolumn{2}{c}{$26.20 \pm 3.44$} & -- \\
\bottomrule
\end{tabular}
\label{tab:resharding_performance}
\end{table}

\subsection{Scalability}
\label{Scalability}

Scaling checkpoint operations to massive, multi-slice environments introduces severe I/O bottlenecks. In this section, we detail saving and loading approaches designed for large training runs in which models replicated across slices.

For saving and loading benchmarks, we use $n$ slices of 64 n2-standard-4 machines with a Llama 3.1 70B model fully
sharded 64 ways and replicated $n$ ways for $n \in \{2, 4, 8, 16, 32\}$. This CPU-only approach to benchmarking exercises the actual resources required for large-scale checkpointing, namely network and distributed storage, without incurring substantial monetary cost from accelerator usage. While device / host transfer throughput cannot be measured with this approach, inter-slice communication between slices uses DCN (data center network) connections regardless of whether TPU or CPU hosts are used.

\subsubsection{Saving}
Traditionally, checkpoint saving in JAX only writes a replicated array shard from the host containing the primary replica, resulting in all write traffic occurring from a single slice, while others sit idle. Orbax introduces the \textbf{Replica-Parallel} feature, which allows all slices to participate in writing a piece of every array. When enabled, each array is partitioned evenly into $n$ segments, corresponding to the $n$ model replicas. Each slice performs device-to-host transfer and write to storage on its designated $1/n$ fraction of the array. As model sizes scale, this approach ensures that the primary slice's bandwidth does not cap the saving throughput. By distributing the I/O workload, parallel saving fully leverages the aggregate network bandwidth available across all participating hosts.

We observe in Table \ref{tab:save_replica_parallel} that the replica-parallel mode significantly improves saving times across all scales. The baseline approach, which saves from a single primary slice, becomes more expensive as the number of replicas grows. Conversely, the replica parallel approach distributes the I/O workload, yielding speedups up to 2.39$\times$ while also reducing the variance in saving times. However, the benefits of the feature are bounded, as subdividing shards too finely over an ever larger number of slices results in an elevated number of queries per second (QPS) for a constant data volume, a traffic pattern that cannot be performantly handled by most storage backends. 

\begin{table}[!htb]
\centering
\caption{Mean checkpoint saving time (seconds) in multi-slice setting.}
\begin{tabular}{crrl}
\toprule
& \multicolumn{2}{c}{Mean Time $\pm$ Std} & \\
\cmidrule(lr){2-3}
Slices & Single-Slice & Replica-Parallel & Speedup \\
\midrule
2 & $ 49.20 \pm 12.28 $ & $ 29.40 \pm 3.00 $ & \textbf{1.67$\times$} \\
4 & $ 49.63 \pm 6.86 $ & $ 25.12 \pm 7.19 $ & \textbf{1.98$\times$} \\
8 & $ 57.46 \pm 20.19 $ & $ 24.47 \pm 6.97 $ & \textbf{2.35$\times$} \\
16 & $ 77.53 \pm 18.88 $ & $ 32.51 \pm 13.88 $ & \textbf{2.39$\times$} \\
32 & $ 131.83 \pm 20.37 $ & $ 77.00 \pm 19.85 $ & \textbf{1.71$\times$} \\
\bottomrule
\end{tabular}

\vspace{1ex}
\label{tab:save_replica_parallel}
\end{table}

We may also expect a reduction in blocking time since each slice need only transfer $1/n$ as much data from devices to hosts. However, our CPU-only benchmarking setup could not observe this behavior.

\subsubsection{Loading}
For large training runs with models replicated across slices, parallel reads from all hosts introduces excessive load on the distributed storage. As the number of hosts scales into the thousands, bandwidth to a distributed storage like GCS can become saturated. The volume of queries per second can itself become a bottleneck even without large data I/O; `lightweight' operations like file listing can become intolerably slow. This negatively impacts training recovery times and overall resource utilization.

Since the bottleneck in this case is primarily duplicate reads from storage, we can potentially reduce the amount of traffic by a factor of \code{n} (the number of model replicas). This is achieved by obtaining shardings for all model weights that are scoped to a single slice of devices, within which a complete replica of the model can be obtained. Loading can then be executed only on the hosts of belonging to the single slice, placed on devices, and broadcast to all other slices using standard JAX APIs. (Note that while we have used the terminology `slice' above, the feature is not tied to physical TPU pods. Users may use the device mesh to designate logical replicas, each of which is a group of devices with a complete replica of the model.)

Using the same hardware and model configuration detailed for the saving experiments (Table \ref{tab:save_replica_parallel}), Table \ref{tab:load_and_broadcast} demonstrates significant improvements in loading time relative to the baseline approach at all sizes except for \code{n=2}, where the overhead of broadcasting dominates the cost of duplicative loading. The baseline approach becomes much more expensive as the number of slices increases, with the broadcasting approach becoming only slightly more expensive due to the increased cost of the broadcast itself. The broadcasting approach also significantly reduces variability at larger scales. 

\begin{table}[!htb]
\centering
\caption{Mean checkpoint load time (seconds) in multi-slice setting.}
\begin{tabular}{crrl}
\toprule
 & \multicolumn{2}{c}{Mean Time $\pm$ Std} & \\
\cmidrule(lr){2-3}
Slices & Broadcast & No Broadcast & Speedup \\
\midrule
2  & $35.31 \pm 1.69$ & $27.88 \pm 2.10$ & 0.79$\times$\textsuperscript{*} \\
4  & $33.33 \pm 2.55$ & $51.77 \pm 3.22$ & \textbf{1.55$\times$} \\
16 & $38.45 \pm 1.45$ & $156.03 \pm 13.68$ & \textbf{4.06$\times$} \\
32 & $46.85 \pm 3.33$ & $231.13 \pm 10.11$ & \textbf{4.93$\times$} \\
\bottomrule
\end{tabular}

\vspace{1ex}
\raggedright
\footnotesize{\textsuperscript{*} \textit{Performance regression where broadcast overhead dominates.}}
\label{tab:load_and_broadcast}
\end{table}

Our simulation approach does suffer from a few limitations, namely that TPU hosts are typically equipped with higher-throughput network interfaces than the machines we used for benchmarking. JAX collectives implementations are also different on CPU and TPU. Due to technical constraints, we were unable to rely on \code{megascale}, the framework of choice for multi-slice training, and needed to rely on the default \code{gloo} implementation, the performance of which is largely unproven. Nevertheless, given that we saw substantial gains even with these limitations, our benchmark may be considered a lower bound for a real scenario incorporating \code{megascale} and actual TPUs. 

\newpage
\bibliography{main}

@misc{geminiteam2025geminifamilyhighlycapable,
      title={Gemini: A Family of Highly Capable Multimodal Models}, 
      author={{Rohan Anil et al.}},
      year={2025},
      eprint={2312.11805},
      archivePrefix={arXiv},
      primaryClass={cs.CL},
      url={https://arxiv.org/abs/2312.11805}, 
}

@misc{roberts2022scalingmodelsdatatextttt5x,
      title={Scaling Up Models and Data with $\texttt{t5x}$ and $\texttt{seqio}$}, 
      author={Adam Roberts and Hyung Won Chung and Anselm Levskaya and Gaurav Mishra and James Bradbury and Daniel Andor and Sharan Narang and Brian Lester and Colin Gaffney and Afroz Mohiuddin and Curtis Hawthorne and Aitor Lewkowycz and Alex Salcianu and Marc van Zee and Jacob Austin and Sebastian Goodman and Livio Baldini Soares and Haitang Hu and Sasha Tsvyashchenko and Aakanksha Chowdhery and Jasmijn Bastings and Jannis Bulian and Xavier Garcia and Jianmo Ni and Andrew Chen and Kathleen Kenealy and Jonathan H. Clark and Stephan Lee and Dan Garrette and James Lee-Thorp and Colin Raffel and Noam Shazeer and Marvin Ritter and Maarten Bosma and Alexandre Passos and Jeremy Maitin-Shepard and Noah Fiedel and Mark Omernick and Brennan Saeta and Ryan Sepassi and Alexander Spiridonov and Joshua Newlan and Andrea Gesmundo},
      year={2022},
      eprint={2203.17189},
      archivePrefix={arXiv},
      primaryClass={cs.LG},
      url={https://arxiv.org/abs/2203.17189}, 
}

@misc{paxml2022github,
  author = {{The Paxml Authors}},
  title = {{Paxml}: A {JAX}-based machine learning framework for training large scale models},
  year = {2022},
  publisher = {GitHub},
  journal = {GitHub repository},
  howpublished = {\url{https://github.com/google/paxml}},
}

@misc{barham2022pathwaysasynchronousdistributeddataflow,
      title={Pathways: Asynchronous Distributed Dataflow for ML}, 
      author={Paul Barham and Aakanksha Chowdhery and Jeff Dean and Sanjay Ghemawat and Steven Hand and Dan Hurt and Michael Isard and Hyeontaek Lim and Ruoming Pang and Sudip Roy and Brennan Saeta and Parker Schuh and Ryan Sepassi and Laurent El Shafey and Chandramohan A. Thekkath and Yonghui Wu},
      year={2022},
      eprint={2203.12533},
      archivePrefix={arXiv},
      primaryClass={cs.DC},
      url={https://arxiv.org/abs/2203.12533}, 
}

@misc{tensorstore2020github,
  author = {{The TensorStore Authors}},
  title = {{TensorStore}: Library for reading and writing large multi-dimensional arrays},
  year = {2020},
  publisher = {GitHub},
  journal = {GitHub repository},
  howpublished = {\url{https://github.com/google/tensorstore}},
}

@misc{jax2018github,
  author = {James Bradbury and Roy Frostig and Peter Hawkins and Matthew James Johnson and Chris Leary and Dougal Maclaurin and George Necula and Adam Paszke and Jake Vander{P}las and Skye Wanderman-{M}ilne and Qiao Zhang},
  title = {{JAX}: composable transformations of {P}ython+{N}um{P}y programs},
  url = {http://github.com/jax-ml/jax},
  year = {2018},
}

@misc{flax2020github,
  author = {Jonathan Heek and Anselm Levskaya and Avital Oliver and Marvin Ritter and Bertrand Rondepierre and Andreas Steiner and Marc van {Z}ee},
  title = {{Flax}: A neural network library and ecosystem for {JAX}},
  url = {http://github.com/google/flax},
  year = {2024},
}

@misc{grain2023github,
  author = {Marvin Ritter and Ihor Indyk and Aayush Singh and Andrew Audibert and Anoosha Seelam and Camelia Hanes and Eric Lau and Jacek Olesiak and Jiyang Kang and Xihui Wu},
  title = {{Grain}: Feeding {JAX} Models},
  url = {http://github.com/google/grain},
  year = {2023},
}

@misc{jax2024colocated,
  title = {Colocated Python},
  author = {{The JAX Authors}},
  year = {2024},
  howpublished = {\url{https://docs.jax.dev/en/latest/notebooks/colocated-python.html}},
  note = {Accessed: 2026-01-13}
}

@misc{jaxstack2024,
  title = {JAX AI Stack},
  author = {{JAX Team}},
  year = {2024},
  howpublished = {\url{https://docs.jaxstack.ai/en/latest/index.html}},
  note = {Accessed: 2026-01-16}
}

@misc{hu2021loralowrankadaptationlarge,
      title={LoRA: Low-Rank Adaptation of Large Language Models}, 
      author={Edward J. Hu and Yelong Shen and Phillip Wallis and Zeyuan Allen-Zhu and Yuanzhi Li and Shean Wang and Lu Wang and Weizhu Chen},
      year={2021},
      eprint={2106.09685},
      archivePrefix={arXiv},
      primaryClass={cs.CL},
      url={https://arxiv.org/abs/2106.09685}, 
}

@misc{wan2025bytecheckpointunifiedcheckpointinglarge,
      title={ByteCheckpoint: A Unified Checkpointing System for Large Foundation Model Development}, 
      author={Borui Wan and Mingji Han and Yiyao Sheng and Yanghua Peng and Haibin Lin and Mofan Zhang and Zhichao Lai and Menghan Yu and Junda Zhang and Zuquan Song and Xin Liu and Chuan Wu},
      year={2025},
      eprint={2407.20143},
      archivePrefix={arXiv},
      primaryClass={cs.AI},
      url={https://arxiv.org/abs/2407.20143}, 
}

@misc{cottier2025risingcoststrainingfrontier,
      title={The rising costs of training frontier AI models}, 
      author={Ben Cottier and Robi Rahman and Loredana Fattorini and Nestor Maslej and Tamay Besiroglu and David Owen},
      year={2025},
      eprint={2405.21015},
      archivePrefix={arXiv},
      primaryClass={cs.CY},
      url={https://arxiv.org/abs/2405.21015}, 
}

@misc{kokolis2025revisitingreliabilitylargescalemachine,
      title={Revisiting Reliability in Large-Scale Machine Learning Research Clusters}, 
      author={Apostolos Kokolis and Michael Kuchnik and John Hoffman and Adithya Kumar and Parth Malani and Faye Ma and Zachary DeVito and Shubho Sengupta and Kalyan Saladi and Carole-Jean Wu},
      year={2025},
      eprint={2410.21680},
      archivePrefix={arXiv},
      primaryClass={cs.DC},
      url={https://arxiv.org/abs/2410.21680}, 
}

@inproceedings{mohan2021checkfreq,
  title = {CheckFreq: Frequent, Fine-Grained {DNN} Checkpointing},
  author = {Mohan, Jayashree and Phanishayee, Amar and Chidambaram, Vijay},
  booktitle = {19th USENIX Conference on File and Storage Technologies (FAST 21)},
  year = {2021},
  pages = {203--216},
  publisher = {USENIX Association},
  month = feb,
  url = {https://www.usenix.org/conference/fast21/presentation/mohan},
}

@misc{maxtext,
  author = {{Google} and {MaxText Contributors}},
  title = {MaxText: A simple, performant and scalable Jax LLM!},
  year = {2024},
  url = {https://github.com/AI-Hypercomputer/maxtext},
  note = {Accessed: 2026-02-10}
}

@manual{pytorch_dcp_2022,
  title  = {torch.distributed.checkpoint},
  author = {{PyTorch Contributors}},
  organization = {PyTorch},
  year   = {2022},
  url    = {https://pytorch.org/docs/stable/distributed.checkpoint.html}
}

@inproceedings{nicolae2020deepfreeze,
  title={DeepFreeze: Towards Scalable Asynchronous Checkpointing of Deep Learning Models},
  author={Nicolae, Bogdan and Li, Jiali and Wozniak, Justin M. and Bosilca, George and Dorier, Matthieu and Cappello, Franck},
  booktitle={2020 20th IEEE/ACM International Symposium on Cluster, Cloud and Internet Computing (CCGRID)},
  pages={172--181},
  year={2020},
  organization={IEEE},
  doi={10.1109/CCGrid49817.2020.00-78}
}

@misc{jin2025adaptivefaulttolerancemechanisms,
      title={Adaptive Fault Tolerance Mechanisms of Large Language Models in Cloud Computing Environments}, 
      author={Yihong Jin and Ze Yang and Xinhe Xu and Yihan Zhang and Shuyang Ji},
      year={2025},
      eprint={2503.12228},
      archivePrefix={arXiv},
      primaryClass={cs.DC},
      url={https://doi.org/10.48550/arXiv.2503.12228}, 
}

@article{kuchaiev2019nemo,
  title={Nemo: a toolkit for building ai applications using neural modules},
  author={Kuchaiev, Oleksii and Li, Jason and Nguyen, Huyen and Hrinchuk, Oleksii and Leary, Ryan and Ginsburg, Boris and Kriman, Samuel and Beliaev, Stanislav and Lavrukhin, Vitaly and Cook, Jack and others},
  journal={arXiv preprint arXiv:1909.09577},
  year={2019}
}

@misc{safetensors2023hf,
  author = {Clémentine Fourrier and Benjamin Bossan and Nicolas Patry and Sylvain Gugger},
  title = {Safetensors: A simple, safe and fast format for storing tensors},
  year = {2023},
  publisher = {GitHub},
  journal = {GitHub repository},
  howpublished = {\url{https://github.com/huggingface/safetensors}},
  commit = {latest}
}

\end{document}